\documentclass[10pt,aps,prl,twocolumn,groupedaddress,longbibliography,
showpacs]{revtex4-1}
\usepackage{epsfig,amsmath,amssymb,MnSymbol,verbatim,bbm}
\usepackage[colorlinks=true,citecolor=blue]{hyperref}

\begin{document}


\title{Thermodynamic Bounds on Precision in Ballistic Multi-Terminal 
Transport}
\author{Kay Brandner\textsuperscript{1}}
\author{Taro Hanazato\textsuperscript{2}}
\author{Keiji Saito\textsuperscript{2}}
\affiliation{\textsuperscript{{{\rm 1}}}Department of Applied Physics,
Aalto University, 00076 Aalto, Finland\\
\textsuperscript{{{\rm 2}}}Department of Physics, Keio University,
3-14-1 Hiyoshi, Yokohama 223-8522, Japan}




\begin{abstract}
For classical ballistic transport in a multi-terminal geometry, we 
derive a universal trade-off relation between total dissipation and 
the precision, at which particles are extracted from individual
reservoirs.
Remarkably, this bound becomes significantly weaker in presence of a
magnetic field breaking time-reversal symmetry.
By working out an explicit model for chiral transport enforced by a
strong magnetic field, we show that our bounds are tight.
Beyond the classical regime, we find that, in quantum systems 
far from equilibrium, correlated exchange of particles makes it
possible to exponentially reduce the thermodynamic cost of precision. 
\end{abstract}


\maketitle

\vbadness=10000
\hbadness=10000

\newcommand{\bzeta}{\boldsymbol{\zeta}}
\newcommand{\bxi}{\boldsymbol{\xi}}
\newcommand{\bB}{\mathbf{B}}
\newcommand{\bs}{\mathbf{s}}
\newcommand{\bS}{\mathbf{S}}
\newcommand{\bT}{\mathbf{T}}
\newcommand{\calT}{\mathcal{T}}
\newcommand{\calS}{\mathcal{S}}
\newcommand{\calF}{\mathcal{F}}
\newcommand{\calV}{\mathcal{V}}
\newcommand{\calD}{\mathcal{D}}
\newcommand{\calE}{\mathcal{E}}
\newcommand{\calQ}{\mathcal{Q}}
\newcommand{\kb}{k_{{{\rm B}}}}
\newcommand{\tr}{{{{\rm tr}}}}

Heisenberg's uncertainty principle is a paradigm example for the 
ubiquitous interplay between fluctuations and precision. 
It entails that the accuracy of simultaneous measurements of 
non-commuting observables is subject to a fundamental lower bound 
arising from intrinsic fluctuations in the underlying quantum states.
Quite remarkably, the precision of non-equilibrium thermodynamic
processes might be restricted through thermal fluctuations in a
similar way: 
Barato and Seifert recently suggested that steady-state biomolecular
process are subject to a universal trade-off between entropy 
production and dispersion in the generated output \cite{Barato2015}. 
Since its discovery, this thermodynamic uncertainty relation has 
triggered significant research efforts. 
A general proof based on methods from large-deviation theory was 
given by Gingrich \emph{et al.} for Markov jump processes satisfying a
local detailed balance condition \cite{Gingrich2016,Gingrich2017}.
Further developments include extensions to finite-time
\cite{Pietzonka2017a,Horowitz2017a} and discrete-time 
\cite{Proesmans2017b} processes, Brownian clocks \cite{Barato2016} and
systems obeying Langevin dynamics \cite{Hyeon2017a,Dechant2017}.

\begin{figure}
\center
\includegraphics[scale=.95]{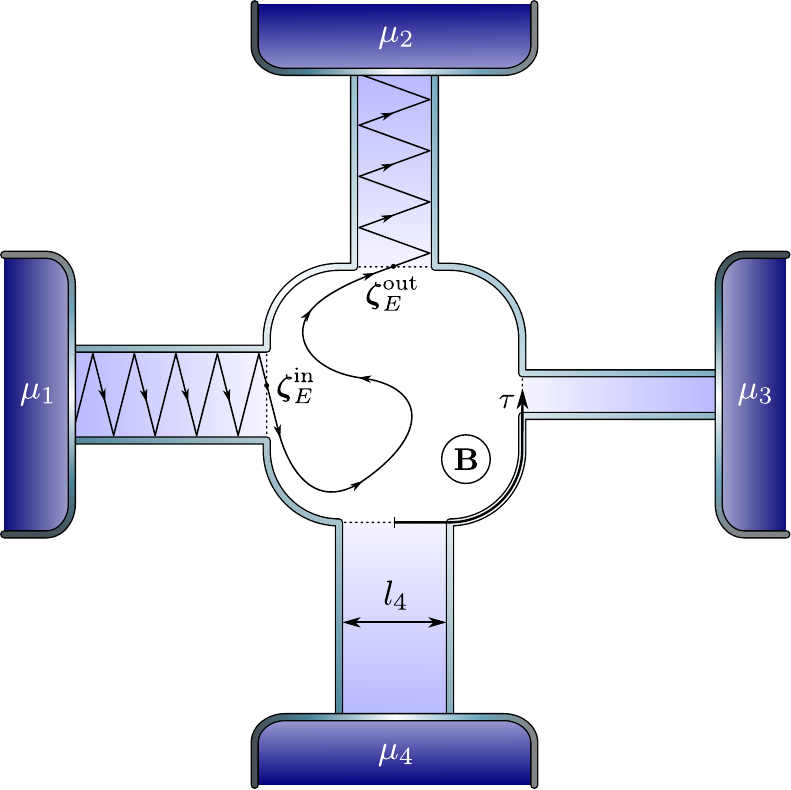}
\caption{Four-terminal setup as an example for multi-terminal 
ballistic transport. 
A two-dimensional target is connected to four reservoirs denoted by
their chemical potentials $\mu_1,\dots,\mu_4$ via perfect leads of
width $l_1,\dots,l_4$. 
The solid line crossing the conductor shows the trajectory of a 
classical particle with energy $E$, which enters the target region at
the point $\bzeta_E^{{{\rm in}}}\equiv (\tau^{{{\rm in}}},
p_\tau^{{{\rm in}}})$ in the reduced phase space and leaves it at 
$\bzeta^{{{\rm out}}}_E\equiv
(\tau^{{{\rm out}}},p_\tau^{{{\rm out}}})$ after being deflected by 
the target potential and the magnetic field $\bB$. 
The coordinate $\tau$ parameterizes the boundary of the target region
and $p_\tau$ denotes the corresponding canonical momentum. 
Since the particle follows Hamiltonian laws of motion, the scattering
map \eqref{ScatMap} is one-to-one. 
Because $\tau$ and $p_\tau$ are canonical variables, the 
Poincar\'e-Cartan theorem implies that this map is also volume
preserving \cite{Ott2009}. 
\label{Fig_Setup}}
\end{figure}

In light of these results, the question arises, whether a fundamental
bound on the precision of thermodynamic processes can be derived from
first principles.
An ideal stage to investigate this problem is provided by ballistic 
conductors, that is, devices, whose dimensions are smaller than the
mean free path of transport carriers.
In such systems, the transfer of particles is governed by reversible
laws of motion, while all irreversible effects are relegated to 
external reservoirs, a mechanism also know as moderate damping
\cite{Humphrey2002,Humphrey2005}.  
This structural simplicity does not only enable the use of physically
transparent models; 
it also leads to a direct link between micro-dynamics and 
thermodynamic observables.
Features such as the inertia of carriers or Lorentz-type
forces, which are not covered by Markov jump processes or overdamped
diffusion, are thereby naturally included. 
These advantages have made ballistic models an important source of
insights on classical \cite{Stark2014,Horvat2009,Casati2008,
Casati2007} and quantum \cite{Brandner2013,Brandner2013a,
Sothmann2014a,Brandner2015,Sanchez2015a} transport mechanisms.
Here, we use this framework to derive a thermodynamic uncertainty
relation for classical ballistic transport, which can be traced
back to elementary properties of Hamiltonian dynamics. 

Scattering theory provides a powerful tool to describe ballistic 
transport in both, the classical and the quantum regime. 
In this approach, the conductor is modeled as a target that is 
connected to $N$ perfect, effectively infinite leads. 
Each lead is attached to a reservoir with fully transparent 
interface injecting thermalized, non-interacting particles. 
Once inside the conductor, the particles follow deterministic 
dynamics until they are absorbed again into one of the reservoirs,
Fig.~\ref{Fig_Setup}. 

On the classical level, the current flowing in the lead $\alpha$ 
towards the target corresponds to a phase-space variable 
$J_\alpha[\bxi_t]$, where the vector $\bxi_t$ contains the positions
and momenta of all particles in the conductor at the time $t$. 
In the steady state, the mean value and fluctuations of this current 
are given by 
\begin{align}
J_\alpha &\equiv \lim_{t\rightarrow\infty}\frac{1}{t}
			   \int_0^t\!\!\!dt'
			   \left\langle J_\alpha[\bxi_{t'}]\right\rangle
               \quad\text{and}
               \label{JSPrim}\\
S_\alpha &\equiv \lim_{t\rightarrow\infty}\frac{1}{t}
                 \int_0^t\!\!\! dt'\!\!\! \int_0^t\!\!\! dt''
                 \left\langle
                 \left(J_\alpha[\bxi_{t'}]-J_\alpha\right)\!
                 \left(J_\alpha[\bxi_{t''}]-J_\alpha\right)
                 \right\rangle,\nonumber 
\end{align}
where the average $\langle\bullet\rangle$ has to be taken over the 
ensemble of trajectories of injected particles \cite{SM}.

Exploiting that the injected particles are statistically independent
and non-interacting, the expressions \eqref{JSPrim} can be made more
explicit.
Focusing on two-dimensional systems from here onwards, to this end,
we decompose the trajectory of a single particle with energy $E$ into
an incoming and an outgoing part connected by the scattering map
\begin{equation}\label{ScatMap}
\mathcal{S}_{E,\bB}:\;\bzeta_E^{{{\rm in}}}
\mapsto \mathcal{S}_{E,\bB}[\bzeta_E^{{{\rm in}}}]
=\bzeta_E^{{{\rm out}}}. 
\end{equation}
The vectors $\bzeta_E^{{{\rm in}}}$ and $\bzeta_E^{{{\rm out}}}$
thereby contain the position and momentum of the particle as it 
enters and leaves the target region and $\bB$ denotes an external
magnetic field applied to the target, Fig.~\ref{Fig_Setup}.  
Using \eqref{ScatMap}, we further introduce the dimensionless 
transmission coefficients 
\begin{equation}\label{TransCoeff}
\calT^{\alpha\beta}_{E,\bB}
= \frac{1}{h}
\int_\beta\! d\bzeta^{{{\rm in}}}_E \int_\alpha\! d\bzeta_E \;
\delta\bigl[\mathcal{S}_{E,\bB}
            [\bzeta^{{{\rm in}}}_E]-\bzeta_E\bigr],
\end{equation}
where indices on integrals imply that the corresponding 
position variable runs only over the boundary between the target 
region and the respective lead, $m$ is the mass of a single particle
and $h$ Planck's constant. 
This definition allows us to compactly rewrite the mean currents and
fluctuations \eqref{JSPrim} as \cite{SM}
\begin{align}
J_\alpha & =\frac{1}{h}\int_0^\infty\!\!\! dE \; \sum_\beta 
              \calT^{\alpha\beta}_{E,\bB}
              \bigl(u^\alpha_E-u^\beta_E\bigr) 
              \quad\text{and}
              \label{JSSinglPart}\\
S_\alpha & =\frac{1}{h}\int_0^\infty\!\!\! dE \sum_{\beta\neq\alpha} 
             \calT^{\alpha\beta}_{E,\bB}
             \bigl(u^\alpha_E + u^\beta_E\bigr).\nonumber 
\end{align}
The chemical potentials $\mu_\alpha$ and temperature $T$ of the
reservoirs enter these expressions via the Maxwell-Boltzmann
distributions
\begin{equation}\label{MBDist}
u^\alpha_E  \equiv \exp\left[-(E-\mu_\alpha)/(\kb T)\right],         
\end{equation}
where $\kb$ denotes Boltzmann's constant. 
Note that the formulas \eqref{JSSinglPart} involve only 
single-particle quantities, while the original definitions 
\eqref{JSPrim} depend on the full phase-space vector $\bxi_t$ of the
many-particle system. 

Maintaining the stationary currents $J_\alpha$ requires a strictly 
positive rate of entropy production 
\cite{Seifert2012,Stark2013,Stark2014}
\begin{equation}\label{EntropyProd}
\sigma  \equiv \kb\sum_\alpha \calF_\alpha J_\alpha
        =\frac{\kb}{h}\int_0^\infty\!\!\! dE \sum_{\alpha\beta} 
         \calT^{\alpha\beta}_{E,\bB}\calF_\alpha 
         \bigl(u^\alpha_E-u^\beta_E\bigr),
\end{equation}
which arises due to heat dissipation in the reservoirs. 
Thus, $\sigma$ can be regarded as the thermodynamic cost of the 
transport process, which is driven by the dimensionless thermodynamic
forces $\calF_\alpha\equiv (\mu_\alpha-\mu)/(\kb T)$ with $\mu$
denoting a reference chemical potential. 

We will now show that this cost puts a universal lower bound on the
relative uncertainty \cite{Barato2015}
\begin{equation}\label{Precision}
\varepsilon_\alpha\equiv S_\alpha/J_\alpha^2
\end{equation}
of each individual current. 
To this end, we consider the quadratic form 
\begin{equation} \label{QuadForm}
A_\alpha\equiv \sigma/\kb+ 2\psi J_\alpha x + \psi S_\alpha x^2,
\end{equation}
where $x,\psi\in\mathbb{R}$. 
For systems without an external magnetic field, $A_\alpha$ can be
written as
\begin{align}
\label{QuadFormSym}
A_\alpha =
\sum_{\beta,\gamma\neq\alpha} \!\! &\calV^{\beta\gamma}
\calD_{\beta\gamma}\bigl(e^{\calD_{\beta\gamma}}-1\bigr)/2\\
+\sum_{\beta\neq\alpha}  &\calV^{\alpha\beta}
\Bigl\{   \bigl(\calD_{\alpha\beta}+ 2\psi x\bigr)
          \bigl(e^{\calD_{\alpha\beta}}-1\bigr)
+\psi x^2 \bigl(e^{\calD_{\alpha\beta}}+1\bigr)\Bigr\}\nonumber
\end{align}
with $\calV^{\alpha\beta}\equiv\int_0^\infty\! dE \; 
\calT^{\alpha\beta}_E u^\beta_E/h\geq 0$ and $\calD_{\alpha\beta}
\equiv\calF_\alpha-\calF_\beta$. 
Here, we used that, at vanishing magnetic field, the transmission
coefficients obey $\calT^{\alpha\beta}_E =\calT^{\beta\alpha}_E$ as a
consequence of time-reversal symmetry \cite{Onsager1931,
Casimir1945,Callen1985,SM}.
Next, we observe that, for any $x$, the second sum in
\eqref{QuadFormSym} is non-negative if $0\leq\psi\leq 2$ 
\footnote{This bound can be verified by minimizing the term inside the
curly brackets in \eqref{QuadFormSym} with respect to $x$ and using
that $y(e^y-1)\geq 2(e^y-1)^2/(e^y+1)$ for any $y\in\mathbb{R}$
\cite{Shiraishi2016b}.}.
Hence, under this condition, the quadratic form $A_\alpha$ is positive
semidefinite, since the first sum in \eqref{QuadFormSym} is generally
non-negative.
Consequently, setting $\psi=2$ in \eqref{QuadForm} and taking the
minimum with respect to $x$ yields 
\begin{equation}\label{URSym}
\sigma\varepsilon_\alpha \geq 2\kb. 
\end{equation}

For systems, where time-reversal symmetry is broken by means of an 
external magnetic field, the transmission coefficients 
$\calT^{\alpha\beta}_{E,\bB}$ are in general not symmetric with
respect to $\alpha$ and $\beta$. 
However, they still fulfill the weaker constraint
$\sum_\beta\calT^{\alpha\beta}_{E,\bB} = 
\sum_\beta\calT^{\beta\alpha}_{E.\bB}$, which follows from the 
volume-preserving property of the scattering map \eqref{ScatMap}
\cite{Stark2013,Stark2014,SM}.
Using this sum rule, the quadratic form \eqref{QuadForm} can be 
expressed as 
\begin{align}
\label{QuadFormAsym}
& A_\alpha = \sum_{\beta\neq\alpha}\sum_{\gamma}
             \calV^{\beta\gamma}_{\bB}
             \bigl(e^{\calD_{\beta\gamma}}-1-\calD_{\beta\gamma}
             \bigr)\\
\hspace*{.229cm}& + \sum_\beta \calV^{\alpha\beta}_{\bB}\Bigl\{
\bigl(1+2\psi x\bigr)\bigl(e^{\calD_{\alpha\beta}}-1\bigr)
+\psi x^2\bigl(e^{\calD_{\alpha\beta}}+1\bigr)
-\calD_{\alpha\beta}\Bigr\},\nonumber
\end{align}
where $\calV^{\alpha\beta}_{\bB}\geq 0$ is defined analogous to
$\calV^{\alpha\beta}$ in \eqref{QuadFormSym}.
Minimizing the term inside the curly brackets shows that the second
sum is \eqref{QuadFormAsym} is non-negative for any $x$ if
\begin{equation}\label{DefPsiAst}
0\leq\psi\leq \min_{y\in\mathbb{R}}
\frac{(1-e^y+ye^y)(e^y+1)}{(e^y-1)^2}
\equiv\psi^\ast\simeq 0.89612.
\end{equation}
Moreover, the first contribution in \eqref{QuadFormAsym}, which does
not depend on $x$, is non-negative due to the convexity of the 
exponential function. 
Hence, by using the same argument as in the derivation of 
\eqref{URSym}, we arrive at 
\begin{equation}\label{URAsym}
\sigma\varepsilon_\alpha\geq\psi^\ast\kb. 
\end{equation}
 
The bounds \eqref{URSym} and \eqref{URAsym} constitute our first main
result. 
Following from elementary microscopic principles, respectively,
time-reversal symmetry and the conservation of phase-space volume, 
they hold for any scattering potential, any number of terminals and
arbitrary far from equilibrium. 
On the macroscopic level, they imply that any increase in the 
precision $1/\varepsilon_\alpha$, at which particles are extracted
from the reservoir $\alpha$, inevitably leads to a proportional
increase of the minimal thermodynamic cost $\sigma$ of the transport
process.
The symmetric bound \eqref{URSym} thereby has exactly the same form as
the recently discovered thermodynamic uncertainty relation for Markov
jump processes \cite{Barato2015,Gingrich2016,Gingrich2017}.
Remarkably, \eqref{URAsym} shows that the minimal cost of precision is
reduced by a factor $\psi^\ast/2$ in systems, where an external
magnetic field breaks the time-reversal symmetry of scattering paths. 

\begin{figure}
\center
\includegraphics[scale=.49]{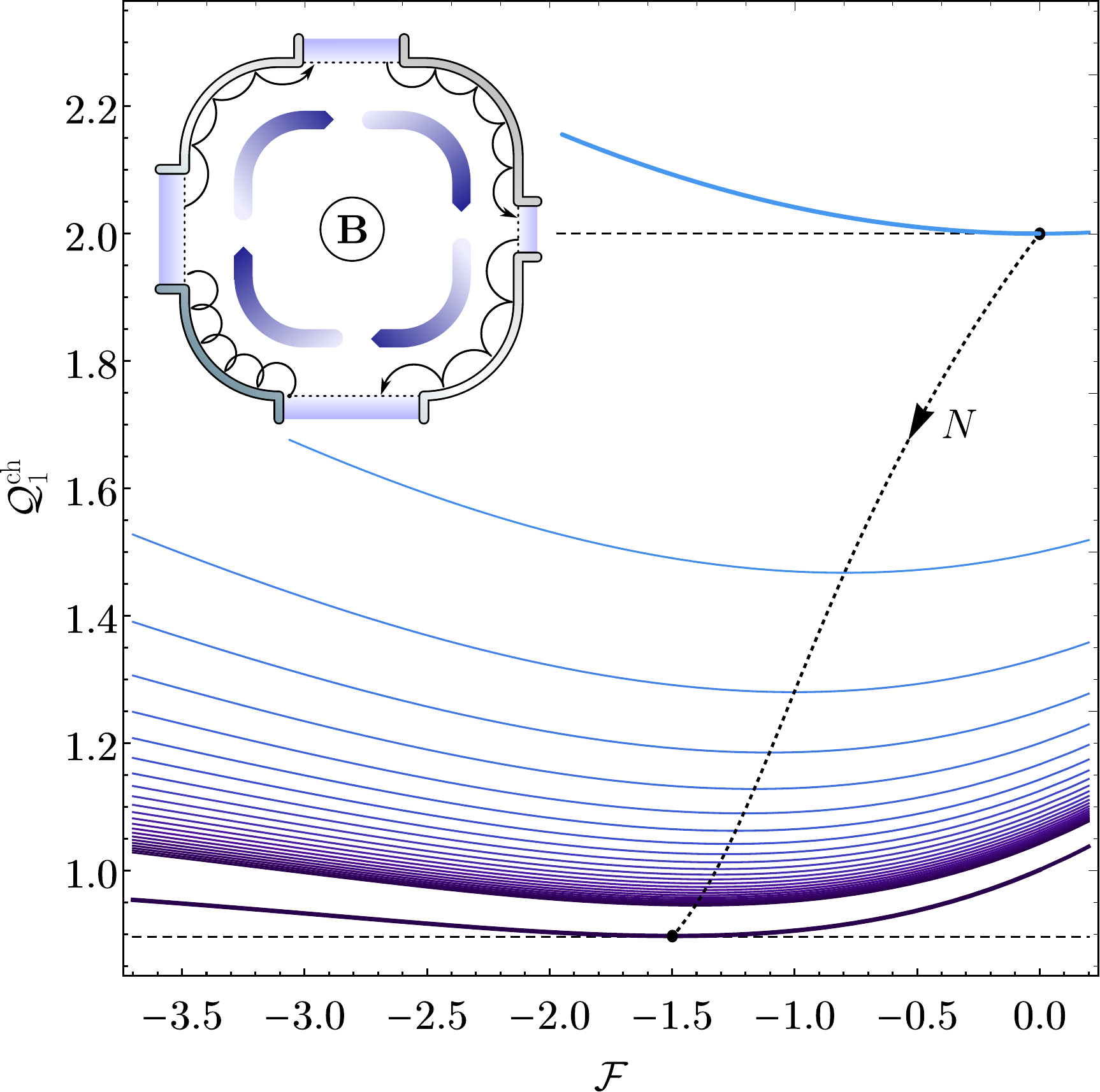}
\caption{\label{Fig_Chira}
Chiral transport. 
Bouncing orbits enforced by a strong magnetic field $\bB$ sustain
clockwise oriented currents between adjacent leads (inset). 
From top to bottom, the blue blue lines show the cost-precision ratio
$\calQ_1$ defined in \eqref{CostPrecChira} as a function of the
rescaled bias increment $\calF$ for systems with $N=2,\dots,25$
terminals and for the limiting case $N\rightarrow\infty$. 
Interpolating between the minima of $\calQ_1$, the dotted line 
crosses the two dashed lines, respectively indicating the bounds 
\eqref{URSym} and \eqref{URAsym}, at $N=2$ and $N\rightarrow\infty$. 
}
\end{figure}

To show that our bounds are tight, we consider an $N$-terminal
conductor with flat target potential.
An external magnetic field $\bB$ forces incoming particles with mass
$m$ and charge $q$ on bouncing orbits along the boundary of the target
region, Fig.~\ref{Fig_Chira}. 
This scattering mechanism is captured by the transmission coefficients
\begin{equation}\label{ChiraTC}
\calT^{\alpha\beta}_{E,\bB} = \sqrt{2mE}\bigl\{
	\pi R_{E,\bB} \delta_{\alpha\beta-1}
   +\left(2l_\alpha-\pi R_{E,\bB}\right)
    \delta_{\alpha\beta}\bigr\}
\end{equation}
with periodic indices $\alpha,\beta=1,\dots,N$ \cite{Stark2014}.
Here, the direction of the magnetic field has been chosen such that
the Larmor circles with radius $R_{E,\bB}\equiv \sqrt{2mE}/(q|\bB|)$
are oriented counterclockwise. 

For a strong magnetic field, the typical Larmor radii are small 
compared to the dimensions of the conductor. 
Under this condition, the transmission coefficients are given by 
\eqref{ChiraTC} throughout the relevant range of energies. 
Due to the asymmetric structure of these coefficients, a chiral steady
state emerges, where currents flow in clockwise direction 
between neighboring reservoirs \cite{Sanchez2015a}.
To generate a net transfer of particles, an external bias has to be 
applied breaking the $N$-fold rotational symmetry of the system. 
For simplicity, here we choose the chemical potentials of the 
reservoirs to increase linearly in steps proportional to $1/N$, that
is we set $\calF_\alpha\equiv \alpha\calF/N$. 
The mean currents and fluctuations can then be evaluated explicitly
by inserting \eqref{ChiraTC} into \eqref{JSSinglPart}.  
Using the abbreviation $\calE\equiv\exp[\calF/N]$, we thus obtain the
expressions 
\begin{align}
\label{CostPrecChira}
\calQ_1 &=\frac{\calF}{N}
\left\{1 +(N-1)\calE^N-\frac{\calE^N-1}{\calE-1}\right\}
\frac{\calE+\calE^N}{(\calE-\calE^N)^2},\\
\calQ_{\alpha} &=\frac{\calF}{N}
\left\{1 +(N-1)\calE^N-\frac{\calE^N-1}{\calE-1}\right\}
\frac{\calE^2+\calE}{\calE^\alpha(\calE-1)^2}
\hspace*{.4678cm}
(\alpha>1)\nonumber
\end{align}
for the dimensionless product $\mathcal{Q}_\alpha\equiv\sigma
\varepsilon_\alpha/\kb$ of total dissipation 
\eqref{EntropyProd} and relative uncertainty \eqref{Precision}. 

In the simplest case $N=2$, the transmission coefficients 
\eqref{ChiraTC} are still symmetric and \eqref{CostPrecChira} reduces
to 
\begin{equation}
\calQ_\alpha = \calF\coth[\calF/4]/2 
                   = 2+ \calF^2/24 + \mathcal{O}\bigl(\calF^4\bigr). 
\end{equation}
Hence, $\calQ_\alpha$ reaches its minimum at $\calF=0$ and the bound
\eqref{URSym} is saturated in the linear response regime. 
As $N$ increases, the minimum of $\calQ_1$ becomes successively
smaller and shifts to negative values of $\calF$, 
Fig.~\ref{Fig_Chira}. 
For large $N$, we obtain the asymptotic expression 
\begin{equation}\label{QNInfChira}
\bigl.
\lim_{N\rightarrow\infty}\calQ_1\bigr|_{\calF<0} 
=\frac{(1-e^\calF+\calF e^\calF)(e^\calF+1)}{(e^\calF-1)^2},
\end{equation}
which should be compared with \eqref{DefPsiAst}. 
In fact, \eqref{QNInfChira} reaches its minimal value $\psi^\ast$ at 
$\calF\simeq -1.49888$. 
This result shows that the cost-precision ratio $\calQ_1$ can come 
arbitrary close to its lower bound \eqref{URAsym} as the number of 
terminals increases. 
By contrast, $\calQ_{\alpha>1}$ asymptotically grows as $N^2$ at any
$\calF\neq 0$. 
This divergence is a consequence of the chiral transmission
coefficients \eqref{ChiraTC} enabling the exchange of particles only
between clockwise-adjacent reservoirs: 
the currents $J_{\alpha>1}$ are effectively driven by the bias
$\calF/N$ and hence vanish as $1/N$, while the fluctuations
$S_{\alpha>1}$ and the total dissipation $\sigma$ stay finite for 
large $N$.

So far, we have shown that precision in classical ballistic transport
requires a minimal thermodynamic cost, which can be substantially 
reduced in systems with broken time-reversal symmetry. 
Although our derivations were performed in a 2-dimensional
setting, it is straightforward to establish \eqref{URSym} and
\eqref{URAsym} also in 1 and 3 dimensions. 
Rather then spelling out the details of this procedure, in the last
part of this article, we develop a perspective beyond the classical
regime. 

For a quantum theory of ballistic transport, the phase-space 
variable $J_\alpha[\bxi_t]$ in \eqref{JSPrim} has to be promoted to
an operator in the Heisenberg picture. 
Replacing classical trajectories with quantum states, the ensemble 
average in \eqref{JSPrim} can then be evaluated using standard
techniques from quantum scattering theory \cite{Gaspard2013a,
Gaspard2015}.
In this formalism, the crucial role of the scattering map
\eqref{ScatMap} is played by the complex scattering matrices 
$\bS_{E,\bB}^{\alpha\beta}$, which connect the amplitudes of incoming
waves in the lead $\beta$ and outgoing waves in the lead $\alpha$,
respectively \cite{Nazarov2009b}. 
For fermionic particles, the mean current is thus obtained as 
\begin{equation}\label{JQuant}
J_\alpha = \frac{1}{h}\int_0^\infty\!\!\! dE \sum_\beta 
\hat{\calT}^{\alpha\beta}_{E,\bB}\bigl(f^\alpha_E -f^\beta_E\bigr). 
\end{equation}
Notably, this expression has the same structure as its classical 
correspondent \eqref{JSSinglPart} with the quantum transmission 
coefficients defined as 
\begin{equation}\label{QuantTC}
\hat{\calT}^{\alpha\beta}_{E,\bB}\equiv
2\;\tr\bigl[\bT_{E,\bB}^{\alpha\beta}\bigr]
         \quad\text{with}\quad
\bT_{E,\bB}^{\alpha\beta}\equiv 
\bS^{\alpha\beta}_{E,\bB}
\bigl(\bS^{\alpha\beta}_{E,\bB}\bigr)^\dagger
\end{equation}
and the Maxwell-Boltzmann distribution \eqref{MBDist} replaced by the
Fermi-Dirac distribution
\begin{equation}\label{FDDist}
f^\alpha_E\equiv 1/\bigl(1+\exp[(E-\mu_\alpha)/(\kb T)]\bigr). 
\end{equation}

The anatomy of current fluctuations in the quantum regime is, 
however, more complicated than in the classical case;
$S_\alpha = S^{{{\rm cl}}}_\alpha-S_\alpha^{{{\rm qu}}}$ 
involves two components \cite{Nazarov2009b}
\begin{align}
\label{SQuant}
S_\alpha^{{{\rm cl}}} &\equiv \frac{1}{h}\!\int_0^\infty\!\!\! dE 
\sum_{\beta\neq\alpha}
\hat{\calT}^{\alpha\beta}_{E,\bB}
\bigl(f^\alpha_E(1-f^\beta_E)+f^\beta_E(1-f^\alpha_E)\bigl),\\
S_\alpha^{{{\rm qu}}} &\equiv \frac{2}{h}\!\int_0^\infty\!\!\! dE
\sum_{\beta\gamma}\tr\bigl[
\bT^{\alpha\beta}_{E,\bB}
\bT^{\alpha\gamma}_{E,\bB}
\bigr]
\bigl(f^\alpha_E-f^\beta_E\bigr)
\bigl(f^\alpha_E-f^\gamma_E\bigr),\nonumber
\end{align}
both of which are non-negative. 
Depending only on single-particle quantities, $S^{{{\rm cl}}}_\alpha$
can be regarded as the quantum analogue of the classical expression
\eqref{JSSinglPart} with additional Pauli-blocking factors accounting
for the exclusion principle. 
By contrast, the contribution $S^{{{\rm qu}}}_\alpha$, which is of
second order in the transmission matrices $\bT^{\alpha\beta}_{E,\bB}$
and hence describes the correlated exchange of two particles, has no 
classical counterpart \cite{Nazarov2009b}.  

The two-component structure \eqref{SQuant} of the current fluctuations
suggests to divide the relative uncertainty $\varepsilon_\alpha=
\varepsilon_\alpha^{{{\rm cl}}}-\varepsilon_\alpha^{{{\rm qu}}}$ into
a quasi-classical part $\varepsilon^{{{\rm cl}}}_\alpha\equiv 
S^{{{\rm cl}}}_\alpha/J_\alpha^2$ and a quantum correction 
$\varepsilon_\alpha^{{{\rm qu}}}\equiv
S^{{{\rm qu}}}_\alpha/J_\alpha^2$. 
By following the lines leading to \eqref{URSym} and \eqref{URAsym} it
is then possible to establish the bounds \cite{SM,Nenciu2007}
\begin{equation}\label{QuantUR}
\sigma\varepsilon^{{{\rm cl}}}_\alpha\geq 2\kb 
\quad\text{and}\quad
\sigma\varepsilon^{{{\rm cl}}}_\alpha\geq \psi^\ast\kb,
\end{equation}
respectively for quantum systems with and without time-reversal
symmetry, where $\sigma=\kb\sum_\alpha \calF_\alpha J_\alpha$. 
As in the classical case, this result follows from the symmetry 
$\hat{\calT}^{\alpha\beta}_{E}=\hat{\calT}^{\beta\alpha}_{E}$ of the 
quantum transmission coefficients \eqref{QuantTC} for
$\bB=0$ and from the sum rules 
$\sum_\beta\hat{\calT}^{\alpha\beta}_{E,\bB}
=\sum_\beta\hat{\calT}^{\beta\alpha}_{E,\bB}$ for $\bB\neq 0$
\cite{Nazarov2009b}.  
It implies in particular that the classical relations \eqref{URSym}
and \eqref{URAsym} are recovered close to equilibrium, i.e., for 
small affinities $\calF_\alpha$, and in the semi-classical regime, 
where the fugacities $\varphi_\alpha\equiv\exp[\mu_\alpha/(\kb T)]$
are small \cite{Callen1985}; in both cases the quantum fluctuations
$S^{{{\rm qu}}}_\alpha$ are negligible. 

In general, however, the quantum corrections
$\varepsilon^{{{\rm qu}}}_\alpha$ will spoil the bounds \eqref{URSym}
and \eqref{URAsym} as the following simple model shows. 
Consider a two-terminal conductor with narrow leads allowing only for
a single open transport channel, i.e., the system is effectively 
1-dimensional and the scattering matrices 
$\bS^{{{\alpha\beta}}}_{E,\bB}$ reduce to complex numbers.  
The target acts as a perfect energy filter, which is fully transparent
in a small window $\Delta$ around the reference chemical potential 
$\mu$ and opaque at all other energies. 
Such filters are standard tools in mesocopic physics
\cite{Yamamoto2015a,Whitney2014,Benenti2017} and can be implemented,
for example, with quantum Hall edge states \cite{Samuelsson2017}.
Setting $\calF_1 \equiv -\calF_2 \equiv \calF/2$ and neglecting 
second-order corrections in $\Delta/(\kb T)$, we obtain 
\begin{equation}
\sigma\varepsilon_\alpha^{{{\rm cl}}} = \kb\calF\coth[\calF/2] 
\;\; \text{and}\;\;
\sigma\varepsilon_\alpha = \kb\calF/\sinh[\calF/2]
\end{equation}
by evaluating \eqref{JQuant} and \eqref{SQuant}. 
Hence, while the product of total dissipation $\sigma$ and 
quasi-classical uncertainty $\varepsilon_\alpha^{{{\rm cl}}}$ is 
bounded by $2\kb$, the full cost-precision ratio $\sigma
\varepsilon_\alpha$ can become arbitrary small. 
Specifically, as $\calF$ becomes large, the current
$J_\alpha$ saturates to a finite value, $\sigma$ grows linearly and 
the fluctuations $S_\alpha$ are exponentially suppressed. 

This example shows that a combination of quantum effects and energy
filtering makes it possible to exponentially reduce the minimal
thermodynamic cost of precision. 
Whether or not this phenomenon can be captured in a generalized 
trade-off relation, where either cost or precision enters 
non-linearly, remains an intriguing question for future research.
Further prospects include the extension of our theory to systems with
temperature gradients or bosonic particles.

Notably, the number $\psi^\ast$, which enters the non-symmetric bounds
\eqref{URAsym} and \eqref{QuantUR}, also appears in a recently found
trade-off relation between power and efficiency of stochastic heat
engines \cite{Shiraishi2016b}. 
These figures are indeed connected with the minimal cost of precision
\cite{Pietzonka2017}. 
Using our approach, it might thus be possible to bound the performance
of ballistic thermoelectric engines, a class of devices that is 
currently subject to active investigations, see for example 
\cite{Horvat2009,Saito2010,Brandner2013,Brandner2013a,Sothmann2014a,
Stark2014,Whitney2014,Brandner2015,Sanchez2015a,Yamamoto2015a,
Samuelsson2017,Benenti2017}. 
At this point, we conclude by stressing that any violation of our
classical bounds constitutes a clear signature of quantum effects. 
Therefore, our work provides a valuable new benchmark to probe 
non-classical transport mechanisms in future theoretical and
experimental studies. 

\newpage
\begin{acknowledgments}
\textbf{Acknowledgments:}
KB acknowledges financial support from the Academy of Finland
(Contract No. 296073) and is affiliated with the Centre of Quantum 
Engineering.
KB thanks P. Pietzonka, P. Burset, M. Moskalets for insightful 
discussions and U. Seifert for longstanding support. 
KS was supported by JSPS Grants-in-Aid for Scientific Research (No. 
JP25103003, JP16H02211 and JP17K05587). 
\end{acknowledgments}

\end{document}



\title{Thermodynamic Bounds on Precision in Ballistic Multi-Terminal 
Transport: Supplemental Material}
\maketitle

\vbadness=10000
\hbadness=10000

\makeatletter
\newcommand{\manuallabel}[2]{\def\@currentlabel{#2}\label{#1}}
\makeatother

\newcommand{\bzeta}{\boldsymbol{\zeta}}
\newcommand{\bxi}{\boldsymbol{\xi}}
\newcommand{\bB}{\mathbf{B}}
\newcommand{\bs}{\mathbf{s}}
\newcommand{\bS}{\mathbf{S}}
\newcommand{\bT}{\mathbf{T}}
\newcommand{\calT}{\mathcal{T}}
\newcommand{\calS}{\mathcal{S}}
\newcommand{\calF}{\mathcal{F}}
\newcommand{\calV}{\mathcal{V}}
\newcommand{\calD}{\mathcal{D}}
\newcommand{\calE}{\mathcal{E}}
\newcommand{\calQ}{\mathcal{Q}}
\newcommand{\calM}{\mathcal{M}}
\newcommand{\calK}{\mathcal{K}}
\newcommand{\calR}{\mathcal{R}}
\newcommand{\calW}{\mathcal{W}}
\newcommand{\calA}{\mathcal{A}}
\newcommand{\kb}{k_{{{\rm B}}}}
\newcommand{\tr}{{{{\rm tr}}}}

\manuallabel{MT_JSClass}{1}
\manuallabel{MT_ClassTransCoeff}{3}
\manuallabel{MT_ScattMap}{2}
\manuallabel{MT_JSClassExpl}{4}
\manuallabel{MT_MaxBoltz}{5}
\manuallabel{MT_AClassSym}{9}
\manuallabel{MT_QuantJ}{18}
\manuallabel{MT_QuantTransCoeff}{19}
\manuallabel{MT_FermiFunct}{20}
\manuallabel{MT_QuantS}{21}
\manuallabel{MT_QuantUncert}{22}

\allowdisplaybreaks

\section{Classical Ballistic Transport}

Starting from the definition Eq.~\ref{MT_JSClass}, we derive the
expressions Eq.~\ref{MT_JSClassExpl} for the mean currents $J_\alpha$
and the current fluctuations $S_\alpha$ in sections A and B. 
In section C, we show that the transmission coefficients defined in 
Eq.~\ref{MT_ClassTransCoeff} obey the relations
\begin{equation}\label{CTF_T}
\calT_{E,\bB}^{\alpha\beta}= \calT^{\beta\alpha}_{E,-\bB}
\quad\text{and}\quad
\sum_\beta\calT_{E,\bB}^{\alpha\beta} = 
\sum_\beta\calT_{E,\bB}^{\beta\alpha}
\end{equation}
as a consequence of time-reversal symmetry and conservation of 
phase-space volume, respectively.

\subsection{Mean Currents}

To implement the classical scattering formalism, we assume that, at 
$t=0$, the target region is empty and each lead is filled with an
ideal gas in equilibrium at the temperature and chemical potential of
the attached reservoir.
This configuration then evolves under Hamiltonian dynamics. 
Since the leads are considered infinite, the long-time limit can be
taken without specifying the mechanism of particle exchange with the
reservoirs \citep{Gaspard2013a}. 
Hence, the ensemble average in Eq.~\ref{MT_JSClass} becomes an average
over the initial number of particles in each lead and the state  
$\bxi_0$ of the system at the initial time $t=0$
\cite{Gaspard2013a}.
Introducing the phase-space variable $n_\alpha[\bxi_t]$, which counts
the number of particles in the lead $\alpha$, we thus have 
\begin{equation}
J_\alpha   \equiv \lim_{t\rightarrow\infty}\frac{1}{t}
           \int_0^t\!\!\! dt' \;
           \langle J_\alpha[\bxi_{t'}]\rangle
          =\lim_{t\rightarrow\infty}\frac{1}{t}
           \langle n_\alpha[\bxi_0]-n_\alpha[\bxi_t]\rangle_{\bxi_0}.
\nonumber       
\end{equation}

To rewrite this expression in terms of single-particle quantities, we
first decompose the phase-space vector $\bxi_t$ into single-particle
components, i.e., 
\begin{equation}
\bxi_t = \bigoplus_{\alpha j}\bxi^{\alpha j}_t.
\nonumber
\end{equation}
Here, the vector $\bxi^{\alpha j}_t$ contains the position and 
momentum coordinates of the particle $j$ initially located in the lead
$\alpha$. 
Second, we note that
\begin{equation}
n_\alpha[\bxi_t] = \sum_{\beta j} \pi_\alpha[\bxi^{\beta j}_t] 
\nonumber
\end{equation}
with $\pi_\alpha[\bxi^{\beta j}_t]=1$ if the particle indexed by
$\beta$  and $j$ is located in the lead $\alpha$ at the time $t$ and 
$\pi_\alpha[\bxi^{\beta j}_t]=0$ otherwise. 
Third, since the particles follow non-interacting, deterministic 
dynamics, the single-particle sate $\bxi^{\alpha j}_t$ at the time 
$t$ is uniquely determined by the corresponding initial state, i.e.,
\begin{equation}
\bxi^{\alpha j}_t = \calM_{t,\bB}[\bxi^{\alpha j}_0],
\nonumber
\end{equation}
where $\calM_{t,\bB}$ is a one-to-one map in the single-particle phase
space depending parametrically on the magnetic field $\bB$.
Hence, using that the gas in each lead is initially described by a
grand canonical ensemble, we obtain
\begin{align}
\label{CTF_J2}
J_\alpha & = \lim_{t\rightarrow\infty}\frac{1}{t}
\biggl\langle \sum_{\beta j} \pi_\alpha[\bxi_0^{\beta j}]
-\pi_\alpha[\calM_{t,\bB}[\bxi^{\beta j}_0]]
 \biggr\rangle_{\bxi_0}\\
& =\lim_{t\rightarrow\infty}\frac{1}{t}
  \sum_\beta \frac{1}{h^2}\int_\beta \!\! d\bxi_0^\beta\; 
  \exp\bigl[-(H[\bxi_0^\beta]-\mu_\beta)/(\kb T)\bigr]\nonumber\\
&\hspace*{4.26cm}
  \bigl(\pi_\alpha[\bxi_0^{\beta}]
 -\pi_\alpha[\calM_{t,\bB}[\bxi^\beta_0]]\bigr).
 \nonumber
\end{align}
Here, $H$ denotes the single-particle Hamiltonian and integrals
with lower index extend over the region in phase space, where the
particle is located in the respective lead. 
For the second expression in \eqref{CTF_J2}, we have used
that all particles originating from the same lead give the same 
contribution to the mean current. 
Therefore, the index $j$ on the single-particle phase-space vector 
can be dropped. 

In a two-dimensional setup, the initial position in phase space of a 
particle approaching the target is determined by four parameters:
its energy $E\in [0,\infty)$, the time $t^{{{\rm in}}} \in [0,\infty)$
the particle needs to reach the target and the coordinates
$\bzeta^{{{\rm in}}}_E\equiv(\tau^{{{\rm in}}},p_\tau^{{{\rm in}}})\in
[0,L]\times[-\sqrt{2mE},\sqrt{2mE}]$, at which the particle enters the
target region. 
Here, $L$ denotes the circumference of the target region. 
For the definition of $\tau$ and $p_\tau$, see Fig.~1 of the main 
text. 
In this parameterization the mean current \eqref{CTF_J2} becomes 
\begin{align}
\label{CTF_J3}
J_\alpha &=\lim_{t\rightarrow\infty} \frac{1}{t}\sum_\beta
\frac{1}{h^2}
\int_0^\infty\!\!\! dE \!\!
\int_0^t\! dt^{{{\rm in}}} \!\!
\int_\beta  d\bzeta^{{{\rm in}}}_E\\
&\hspace*{.37cm}\exp\bigl[-(E-\mu_\beta)/(\kb T)\bigr]
\bigl(\delta_{\alpha\beta}-
\pi_\alpha[\calK_{t,\bB}
[E,t^{{{\rm in}}},\bzeta^{{{\rm in}}}_E]]\bigr),
\nonumber
\end{align} 
where $\calK_{t,\bB}:(E,t^{{{\rm in}}},\bzeta^{{{\rm in}}}_E)\mapsto 
\bxi_t^\alpha$ denotes a one-to-one mapping from the initial
condition specified by the parameters $E,t^{{{\rm in}}}$ and
$\bzeta^{{{\rm in}}}_E$ to the single-particle state $\bxi^\alpha_t$
evolving from it.
Note that this map is volume preserving for any $t\geq 0$ by virtue 
of the Poincar\'e-Cartan theorem \cite{Ott2009}. 

To carry out the long-time limit, we observe that, for large $t$,
the typical time particles spend inside the target region becomes
negligible. 
Therefore, we can replace $\calK_{t,\bB}$ with
$\calK_{\infty,\bB}$ in \eqref{CTF_J3}.
Invoking the relation 
\begin{equation}
\pi_\alpha[\calK_{\infty,\bB}[E,t^{{{\rm in}}},\bzeta^{{{\rm in}}}_E]]
=\int_\alpha \!\! d\bzeta_E \; 
 \delta\bigl[ \calS_{E,\bB}[\bzeta^{{{\rm in}}}_E]-\bzeta_E\bigr]
\nonumber
\end{equation}
with the scattering map $\calS_{E,\bB}$ defined in
Eq.~\ref{MT_ScattMap} and the definitions Eq.~\ref{MT_ClassTransCoeff}
and Eq.~\ref{MT_MaxBoltz}, we thus arrive at 
\begin{align}
\label{CTF_J4}
J_\alpha &=\frac{1}{h}\int_0^\infty \!\!\! dE \; \sum_\beta 
u^\beta_E \bigl( 2 \delta_{\alpha\beta}
l_\alpha \sqrt{2mE}/h -\calT^{\alpha\beta}_{E,\bB}\bigr)\\
&=\frac{1}{h}\int_0^\infty \!\!\! dE \; \sum_\beta 
\calT^{\alpha\beta}_{E,\bB}\bigl(u_E^\alpha-u_E^\beta\bigr). 
\nonumber
\end{align}
The second line thereby follows from the sum rule 
$\sum_\beta \calT^{\alpha\beta}_{E,\bB} = 
2l_\alpha\sqrt{2mE}/h$, which we derive in section C.

\subsection{Current Fluctuations}

The fluctuations $S_\alpha$ can be computed along the same lines as the
mean currents. 
First, using the notation of the previous section, we have 
\begin{align}
S_\alpha & = \lim_{t\rightarrow\infty}\frac{1}{t}
\int_0^t\!\!\! dt' \! \int_0^t\!\!\! dt'' \;
\langle(J_\alpha[\bxi_{t'}]-J_\alpha)(J_\alpha[\bxi_{t''}]-J_\alpha)
\rangle\nonumber\\
& =\lim_{t\rightarrow\infty}\frac{1}{t}\Bigl\{ \!
\langle \! (n_\alpha[\bxi_0]-n_\alpha[\bxi_t])^2\rangle_{\bxi_0}
-\langle n_\alpha[\bxi_0]-n_\alpha[\bxi_t]\rangle_{\bxi_0}^2
\!\Bigr\}\nonumber\\
&= \lim_{t\rightarrow\infty}\frac{1}{t}\biggl\{
\Bigl\langle \Bigl(\sum_{\beta j} 
\pi_\alpha[\bxi^{\beta j}_0] - \pi_\alpha[\calM_{t,\bB}
[\bxi^{\beta j}]]
\Bigl)^2 \Bigr\rangle_{\bxi_0}\nonumber\\
&\hspace*{2.72cm}
-\Bigl\langle\sum_{\beta j} \pi_\alpha[\bxi^{\beta j}_0]
- \pi_\alpha[\calM_{t,\bB}[\bxi^{\beta j}]]
\Bigr\rangle_{\bxi_0}^2\biggr\}.
\nonumber
\end{align}
Second, evaluating the grand-canonical average yields 
\begin{align}
S_\alpha &= \lim_{t\rightarrow\infty}\frac{1}{t}\sum_\beta
\frac{1}{h^2} \int_\beta \!\! d\bxi^\beta_0 \;
\exp\bigl[-(H[\bxi_0^\beta]-\mu_\beta)/(\kb T)\bigr]\nonumber\\
& \hspace*{4.36cm} \bigl( \pi^\alpha[\bxi^\beta_0] 
       - \pi^\alpha[\calM_t[\bxi^\beta_0]]\bigr)^2.
\nonumber
\end{align}
Third, using that $\pi_\alpha^2 = \pi_\alpha$ and repeating the steps
leading from \eqref{CTF_J2} to \eqref{CTF_J4} gives
\begin{align}
S_\alpha &= \frac{1}{h}\int_0^\infty \!\!\! dE  \sum_\beta
u_E^\beta\bigl\{ 2\delta_{\alpha\beta} l_\alpha\sqrt{2mE}/h
-2(\delta_{\alpha\beta}-1)\calT^{\alpha\beta}_{E,\bB}\bigr\}
\nonumber\\
&= \frac{1}{h}\int_0^\infty \!\!\! dE  \sum_\beta
(1-\delta_{\alpha\beta})\calT_{E,\bB}^{\alpha\beta}
\bigl(u^\alpha_E+ u^\beta_E\bigr)\nonumber\\
&= \frac{1}{h}\int_0^\infty \!\!\! dE  \sum_{\beta\neq\alpha}
\calT_{E,\bB}^{\alpha\beta}\bigl(u^\alpha_E+ u^\beta_E\bigr),
\nonumber
\end{align}
where the second line follows from the sum rule $\sum_\beta
\calT^{\alpha\beta}_{E,\bB}=2l_\alpha\sqrt{2mE}/h$. 

\subsection{Transmission Coefficients}

The symmetries \eqref{CTF_T} of the classical transmission 
coefficients result from the properties of the scattering map 
\begin{equation}\label{CTF_TS}
\calS_{E,\bB}:\bzeta^{{{\rm in}}}_E\mapsto
\calS_{E,\bB}[\bzeta^{{{\rm in}}}_E]=\bzeta^{{{\rm out}}}_E
\end{equation}
connecting the reduced phase-space coordinates, at which a particle
with energy $E$ enters and leaves the target region. 
First, we have
\begin{equation}\label{CTF_TSRev}
\calS_{E,-\bB}[\calR[\bzeta^{{{\rm out}}}_E]]
=\calR[\bzeta^{{{\rm in }}}_E]
\end{equation}
with $\calR:\bzeta_E\mapsto\calR[\bzeta_E]$ denoting the time-reversal
map, which switches the sign of momentum coordinates while leaving
spatial coordinates unchanged.       
Thus, a mirror particle injected at $\tau^{{{\rm out}}}$ with energy
$E$ and transversal momentum $-p_\tau^{{{\rm out}}}$, under the 
reversed magnetic field $-\bB$, retraces the trajectory of the original 
particle and leaves the target region at $\tau^{{{\rm in}}}$ with
transversal momentum $-p_\tau^{{{\rm in}}}$. 
Consequently, it follows that
\begin{align}
\nonumber
\calT^{\beta\alpha}_{E,-\bB} &= \frac{1}{h}
\int_\alpha\!\! d\bzeta^{{{\rm in}}}_E\int_\beta \!\! d\bzeta_E \;
\delta\bigl[\calS_{E,-\bB}[\bzeta^{{{\rm in}}}_E]-\bzeta_E\bigr]\\
&= \frac{1}{h}
\int_\alpha\!\! d\bzeta^{{{\rm in}}}_E\int_\beta \!\!d\bzeta_E \;
\delta\bigl[\calR[\bzeta^{{{\rm in}}}_E]
-\calS_{E,-\bB}^{-1}[\calR[\bzeta_E]\bigr]\nonumber\\
&= \frac{1}{h}
\int_\alpha\!\! d\bzeta^{{{\rm in}}}_E\int_\beta \!\!d\bzeta_E \;
\delta\bigl[\calR[\bzeta^{{{\rm in}}}_E]
-\calR[\calS_{E,\bB}[\bzeta_E]]\bigr]\nonumber\\
&= \frac{1}{h} \int_\beta \!\! d\bzeta^{{{\rm in}}}_E
\int_\alpha\!\! d\bzeta_E \;
\delta\bigl[\calS_{E,\bB}[\bzeta_E^{{{\rm in}}}]-\bzeta_E\bigr]
=\calT^{\alpha\beta}_{E,\bB}.
\nonumber
\end{align}
Here, we used in the first step that, for Hamiltonian dynamics, the
scattering map $\calS_{E,-\bB}$ is invertible and volume-preserving
due to the Poinca\'e-Cartan theorem \cite{Ott2009}. 
Furthermore, we exploited that the domain of the reduced phase-space
integrals is invariant under the time-reversal map $\calR$.
In the second step, we applied \eqref{CTF_TS} and \eqref{CTF_TSRev}.
Finally, in the third step, we relabeled the integration variables.

To derive the sum rules \eqref{CTF_TS}, we first note that 
\begin{align}
\label{CTF_TSR1}
\sum_\beta \calT_{E,\bB}^{\beta\alpha} &=\frac{1}{h}\sum_\beta
\int_\alpha \!\! d\tau^{{{\rm in}}} \!\! 
\int_{-\sqrt{2mE}}^{\sqrt{2mE}} dp^{{{\rm in}}}_\tau \!\! 
\int_\beta \!\! d\tau \!\! 
\int_{-\sqrt{2mE}}^{\sqrt{2mE}} dp_\tau\\
& \hspace*{1.53cm}
\delta\bigl[\tau^{{{\rm out}}}[\tau^{{{\rm in}}},p_\tau^{{{\rm in}}}]
             -\tau\bigr]
\delta\bigl[p^{{{\rm out}}}_\tau[\tau^{{{\rm in}}},p_\tau^{{{\rm in}}}]
             - p_\tau\bigr] \nonumber\\
& = 2l_\alpha \sqrt{2mE}/h, \nonumber
\end{align}
where $l_\alpha$ is the width of the lead $\alpha$. 
Second, using again that the scattering map is invertible and 
volume-preserving, we have 
\begin{align}
\label{CTF_TSR2}
\sum_\beta \calT^{\alpha\beta}_{E,\bB} &= \frac{1}{h}\sum_\beta 
\int_\beta \!\! d\bzeta^{{{\rm in}}}_E 
\int_\alpha \!\! d\bzeta_E \; 
\delta\bigl[\bzeta^{{{\rm in}}}_E-\calS^{-1}_{E,\bB}[\bzeta_E]\bigr]\\
&= 2 l_\alpha\sqrt{2mE}/h.
\nonumber
\end{align}
Comparing \eqref{CTF_TSR1} and \eqref{CTF_TSR2} yields the sum rules 
\eqref{CTF_T}.

\section{Quantum Ballistic Transport}

We derive the bounds Eq.~\ref{MT_QuantUncert} on the quasi-classical
part of the relative uncertainty, $\varepsilon^{{{\rm cl}}}_\alpha
\equiv S_\alpha^{{{\rm cl}}}/J_\alpha^2$.
To this end, we first note that, for $\bB=0$, the quantum transmission
coefficients Eq.~\ref{MT_QuantTransCoeff} are symmetric, i.e., 
$\hat{\calT}^{\alpha\beta}_E =\hat{\calT}^{\beta\alpha}_E$
\cite{Nazarov2009b}. 
Hence, using the expression Eq.~\ref{MT_QuantJ} for the mean currents
$J_\alpha$, the total rate of entropy production can be written as 
\begin{equation}\label{QTF_QFSym}
\sigma=\kb\sum_\alpha \calF_\alpha J_\alpha 
=\frac{\kb}{2h}\int_0^\infty\!\!\! dE\sum_{\alpha\beta}
\hat{\calT}^{\alpha\beta}_E
\bigl(\calF_\alpha-\calF_\beta\bigr)
\bigl(f^\alpha_E - f^\beta_E\bigr).
\nonumber
\end{equation}
Together with Eq.~\eqref{MT_QuantS}, this structure implies  
\begin{align}
\label{QTF_QFSym1}
A_\alpha^{{{\rm cl}}} &= \sigma/\kb +2\psi J_\alpha x 
                        +\psi S^{{{\rm cl}}}_\alpha x^2\\
&=\sum_{\beta,\gamma\neq \alpha}\calW^{\beta\gamma}\calD_{\beta\gamma}
        \bigl(e^{\calD_{\beta\gamma}}-1\bigr)/2\nonumber\\
& +\sum_{\beta\neq\alpha}\calW^{\alpha\beta}\Bigl\{
  \bigl(\calD_{\alpha\beta}+2\psi x\bigr)\bigl(e^{\calD_{\alpha\beta}}
  -1\bigr) + \psi x^2 \bigl(e^{\calD_{\alpha\beta}}+1\bigr)\Bigr\},
\nonumber
\end{align}
where $\calD_{\alpha\beta}=\calF_\alpha-\calF_\beta$ as in the main
text. 
Furthermore, the coefficients 
$\calW^{\alpha\beta}\equiv\int_0^\infty dE\; 
\hat{T}^{\alpha\beta}_E(1-f^\alpha_E)f^\beta_E/h\geq 0$ are 
non-negative, since Fermi functions $f^\alpha_E$ defined in 
Eq.~\ref{MT_FermiFunct} are bounded between $0$ and $1$. 
Expression \eqref{QTF_QFSym1} has the same structure as its
classical counterpart Eq.~\ref{MT_AClassSym}. 
It follows that $A^{{{\rm cl}}}_\alpha\geq 0$ for any $x$ provided 
that $0\leq\psi\leq 2$. 
Thus, we end up with
$\sigma\varepsilon^{{{\rm cl}}}_\alpha\geq 2 \kb$.

For $\bB\neq 0$, the sum rules
$\sum_\beta\hat{\calT}^{\alpha\beta}_{E,\bB}
=\sum_\beta\hat{\calT}^{\beta\alpha}_{E,\bB}$ 
\cite{Nazarov2009b}
can be used to express
the total rate of entropy production as
\begin{align}
\nonumber
&\sigma = \kb \sum_\alpha \calF_\alpha J_\alpha\\
       &= \frac{\kb}{h}\int_0^\infty\!\!\! dE
          \sum_{\alpha\beta}\hat{\calT}^{\alpha\beta}_{E,\bB}
          \Bigl\{f[\nu^\beta_E]\bigl(\nu^\beta_E-\nu^\alpha_E\bigr)
                -g[\nu^\beta_E]+g[\nu^\alpha_E]\Bigr\},
\nonumber
\end{align}
where $\nu^\alpha_E\equiv (\mu_\alpha-E)/(\kb T)$,
\begin{equation}
\nonumber
f[y] \equiv 1/\bigl(1+\exp[-y]\bigr)\quad\text{and}\quad
g[y] \equiv y- \ln[f[y]]. 
\end{equation}
Hence, we have 
\begin{align}
\label{QTF_QFAsym}
&A^{{{\rm cl}}}_\alpha= \sigma/\kb +2\psi J_\alpha x 
                       +\psi S^{{{\rm cl}}}_\alpha x^2\\
&= \frac{1}{h}\int_0^\infty\!\!\!\!dE\!
   \sum_{\beta\neq\alpha}\sum_\gamma\hat{\calT}^{\beta\gamma}_{E,\bB}
   \Bigl\{
      f[\nu_E^\gamma]\bigl(\nu^\gamma_E-\nu^\beta_E\bigr)
      -g[\nu^\gamma_E]+g[\nu^\beta_E]
   \Bigr\}\nonumber\\
&+ \frac{1}{h}\int_0^\infty\!\!\!\!dE\sum_\beta\hat{\calT}_{E,\bB}
   \Bigl\{
      f[\nu^\beta_E]\bigl(\nu^\beta_E-\nu^\alpha_E\bigr)
      -g[\nu^\beta_E]+g[\nu^\alpha_E]
   \nonumber\\
&\hspace*{2.64cm}
   +2\psi x\bigl(f[\nu^\alpha_E]-f[\nu^\beta_E]\bigr)
   \nonumber\\
&\hspace*{2.64cm}
   +\psi x^2\bigl(f[\nu^\alpha_E] +f[\nu^\beta_E]
                  -2f[\nu^\alpha_E]f[\nu^\beta_E]\bigr)\Bigr\}.
   \nonumber
\end{align}
Here, the first integral, which does not depend on $x$, is 
non-negative, since the function $g[y]$ is convex and 
$\partial_y g[y]=f[y]$, for details see \cite{Nenciu2007}. 
Therefore, the quadratic form \eqref{QTF_QFAsym} is positive
semidefinite if
\begin{equation}\label{QTF_Min}
0\leq\psi\leq \min_{y,z\geq 0}
\frac{w[y,z](y^2+1)}{(y^2-1)^2}, 
\end{equation}
where 
\begin{equation}
w[y,z] \equiv (y+z)\Bigl\{(y+1/z)\ln\bigl[(y+z)/(1/y+z)\bigr]
-2\ln[y]/z\Bigr\}.
\nonumber 
\end{equation}
This condition can be derived by minimizing the term inside the curly
brackets in the second integral in \eqref{QTF_QFAsym} with respect to
$x$ and expressing the result in the variables 
$y_{\beta\alpha}\equiv\exp\bigl[(\nu^\beta_E-\nu^\alpha_E)/2\bigr]$
and 
$z_E^{\alpha\beta}\equiv \exp\bigl[(\nu^\alpha_E+\nu^\beta_E)/2
\bigr]$.

To bound the minimum \eqref{QTF_Min}, we proceed as follows. 
First, we compute 
\begin{align}
w'[y,z] &= q[y,z]/z^2,\nonumber\\
q[y,z]  &\equiv z(1-y^2) +2y\ln[y]\nonumber\\
&\hspace*{2.54cm} +y\bigl(z^2-1\bigr)
\ln\bigl[(y+z)/(1/y+z)\bigr],\nonumber\\
q'[y,z] &\equiv \bigl\{yz\bigl(1-y^2\bigr)/\bigl(1+yz\bigr)\!\bigr\}
\bigl\{1+u+2u\ln[u]/(1-u)\!\bigr\},\nonumber
\end{align}
where $u\equiv(1/y+z)/(y+z)$ and primes indicate derivatives with
respect to $z$. 
Second, we note that 
\begin{align}
\nonumber
1+u+2u\ln[u]/(1-u)\geq 0 \quad\text{for}\quad u\geq 0.
\end{align}
This relation can be easily proven by using that the logarithm is a
concave function. 
Since $y,z\geq 0$, it implies that 
$q'[y,z]\geq 0$ for $y\leq 1$ and $q'[y,z]\leq 0$ for $y>0$. 
Consequently, since $q[y,z=0]=0$, we have 
\begin{equation}
\nonumber
q[y,z]\geq 0 \;\;\text{for}\;\; y\leq 1\;\;\text{and}\;\;
q[y,z]\leq 0 \;\;\text{for}\;\; y>1
\end{equation}
and the same inequalities hold for $w'[y,z]$. 
Thus, the function $w[y,z]$ is monotonically increasing in $z$ for 
$y\leq 1$ and monotonically decreasing for $y>1$. 

Consequently, it follows that 
\begin{align}
\nonumber
w[y,z] &\geq \lim_{z\rightarrow 0} w[y,z]
=1-y^2+2y^2\ln[y] \;\; \text{for}\;\; y\leq 1 \;\;\text{and}\\
w[y,z] &\geq \lim_{z\rightarrow\infty} w[y,z]
= y^2 -1 -2\ln[y]\;\; \text{for} \;\; y>1. \nonumber
\end{align}
Combining these relations, we arrive at 
\begin{align}
\nonumber
\min_{y,z\geq 0} \frac{w[y,z](y^2+1)}{(y^2-1)^2}
&\geq \min_{0\leq y\leq 1}\frac{(1-y^2+2y^2\ln[y])(y^2+1)}{(y^2-1)^2}
\\
&= \min_{y\leq 0}\frac{(1-e^y+ye^y)(e^y+1)}{(e^y-1)^2} =\psi^\ast,
\nonumber
\end{align}
where the minimum in the second line is attained for $y^\ast\simeq 
-1.49888$. 
Thus, the condition \eqref{QTF_Min} is fulfilled for any $0\leq\psi
\leq \psi^\ast$. 
Finally, setting $\psi=\psi^\ast$ in the first line of
\eqref{QTF_QFAsym} and taking the minimum with respect to $x$ yields
$\sigma\varepsilon^{{{\rm cl}}}\geq \psi^\ast k_B$.